\documentclass[a4,prl,showpacs,showkeys, sort&compress, floatfix, square]{revtex4}

\usepackage[dvips]{graphicx}

\usepackage{bm}

\renewcommand{\deg}{^\circ}

\begin{document}

\title{Helical anisotropy and  magnetoimpedance of CoFeSiB wires under torsional stress}

\author{C. Tannous, J. Gieraltowski and R. Valenzuela\dag}
\affiliation{Laboratoire de Magn\'{e}tisme de Bretagne,  CNRS FRE-2697, BP: 809 Brest Cedex,  29285 France \\
\dag Instituto de Investigaciones en Materiales, Universidad Nacional Aut\'{o}noma de Mexico,
 P.O. Box 70-360, Coyoacan, Mexico D.F., 04510, Mexico}

\begin{abstract}
Recent measurements of the magnetoimpedance (at a fixed frequency of 1 MHz) of cobalt-rich wires 
subjected to torsion stress show an asymmetry as a function of torsion angle stemming  
from residual anisotropies induced during wire fabrication.
We interpret these measurements with a simple model based on the competition between a circumferential
magnetic anisotropy and another one induced by torsional and residual stress. This allows extraction of the physical
parameters of the wire and explains the positive and negative torsion cases. The agreement between
theoretical and experimental results provides a firm support for the model describing the behaviour 
of the anisotropy field versus static magnetic field for all torsion angles.
\end{abstract}

\pacs {75.50.Kj;   72.15.Gd;   75.30.Gw;  75.80.+q}

\keywords {Amorphous magnetic wires; Giant magnetoimpedance;  Magnetic anisotropy; Magnetostriction}

\maketitle

\section{Introduction}

The giant magneto-impedance effect (GMI) in amorphous ribbons, wires and thin films 
has become a topic of growing interest for a wide variety of prospective 
applications in storage information technology and sensors possessing high 
sensivity and fast response \cite{mohri, panina}. 

Magneto-impedance effect (MI) consists in the change of impedance introduced by 
a low-amplitude ac current $I_{ac}$ flowing through a magnetic conductor 
under application of a static magnetic field $H_{dc}$ (usually parallel 
to the direction of the ac current). The origin of this behavior is related 
to the relative magnetic  permeability $\mu_{r}$ and the direction 
of the magnetic anisotropy field $H_{k}$.  When MI measurements are performed 
on wires, injecting an ac current in the wire
and applying a dc magnetic field $H_{dc}$ along the wire axis, one probes the
wire rotational permeability wire that controls the behaviour of the MI.

Ordinarily, the MI curve is symmetric with respect to the dc magnetic field $H_{dc}$ 
with a single (at $H_{dc}$=0) or a double peak (when $H_{dc}= \pm H_{max}$), the anisotropy
field of the wire being considered as the value of $H_{max}$. Applying a torsion stress to the
wire alters its MI behaviour versus field $H_{dc}$ in a way such that one observes, 
for instance, a single peak only for positive stress and a double peak for negative stress.

Recent work by Betancourt and Valenzuela (BV) \cite{BV}  tackled  MI measurements
of CoFeBSi wires subjected to torsion stress \cite{helical}, at a fixed frequency of 1 MHz,
obtaining the aforementioned asymmetry \cite{blanco, Ryu} of the MI profile with 
respect to torsion stress.  Interpreting the asymmetry in terms of residual anisotropies
induced during wire fabrication \cite{BV} leads to an effect that respectively
counterbalances or enhances the wire circumferential anisotropy with positive or negative torsional stress.

The wires were prepared by the in-rotating water-quenching technique and have the nominal 
composition (Co$_{94}$Fe$_{6}$)$_{72.5}$Si$_{12.5}$B$_{15}$. Their typical dimensions are 10 cm in 
length and 120 $\mu$m for diameter. They possess a characteristic domain structure dictated by the sign
of the magnetostriction coefficient $\lambda_{s}^{R}$ which is expected to have a small 
value ($|\lambda_{s}^{R}| \sim 10^{-7}$ in these Cobalt rich alloys). When 
$\lambda_{s}^{R}$ is negative, an axially oriented magnetization core exists 
surrounded by circular domains \cite{mohri, panina}, whereas in 
the opposite case, radially oriented magnetization states exist such as those observed
in Fe rich alloys \cite{Ryu}. 
After performing several measurements such as Barkhausen jump and magnetization reversal 
measurements \cite{BV}, BV favor the negative sign for the $\lambda_{s}^{R}$ coefficient and
presence of helical anisotropy \cite{helical}.

In this paper, a simple model for the anisotropy field, based on the competition between 
an existing circumferential anisotropy and a torsion-induced one with presence of 
residual torsional stress, is proposed. This model allows the extraction 
of the relevant physical parameters of the wire and explains the observed behavior in the cited
work \cite{BV} as a function of the static magnetic field for all torsion angles.

\section{Theory}
\begin{figure}[!h]
\begin{center}
\scalebox{0.5}{\includegraphics[angle=0]{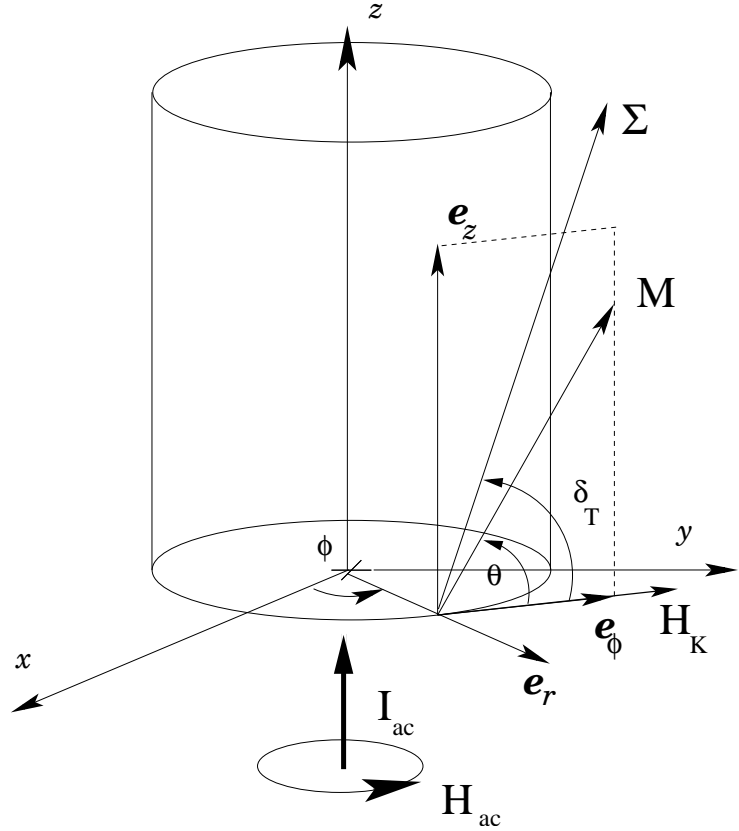}}
\end{center}
\caption{Geometry of the anisotropy and stress-induced easy axis.  The circumferential anisotropy 
axis induced by fabrication is along  $\bf e_{\phi}$  and the torsion stress-induced easy axis  $\bf \Sigma$
makes an angle $\delta_T$ with $\bf e_{\phi}$. The zero-stress circular anisotropy field  $H_{K}$ is 
along $\bf e_{\phi}$.
In the helical anisotropy case, the magnetization $\bf M$ lies in the $\bf e_{\phi},
\bf e_{z}$ plane making a non zero angle $\theta$ with the circumferential basis vector $\bf e_{\phi}$, 
otherwise, both the magnetization $\bf M$ and anisotropy axis lie in the $\bf e_{r}, \bf e_{\phi}$ plane.}  
\label{fig1}
\end{figure}

In order to reveal the residual stress within the wires, positive and negative
torsion stress \cite{helical} are applied and the effect is analysed with MI response.
Following the work of Favieres {\it et al.} \cite{Favieres} on cylindrical CoP amorphous
multilayers electrodeposited on copper wires,
we use a stress-induced easy axis making an arbitrary angle  $\delta_T$ \cite{helical}
with the circumferential direction (see fig.~\ref{fig1}). We extend
this model to include the presence of residual torsion stress induced by fabrication.
The total  magnetoelastic energy in presence of applied and  residual torsional stress ($\sigma_r$) is:

\begin{equation}
E= K_c \sin^{2}(\theta) + \frac{3}{2}\lambda_{s}^{R} (\sigma + \sigma_r) \sin^{2}(\delta_T - \theta)
\end{equation}

where $K_c$ is the fabrication induced circumferential anisotropy of the wire and $\lambda^{R}_{s}$  the rotational
saturation magnetostriction constant \cite{constant} (see fig.1). We are in the linear case where the magnetoelastic energy 
is proportional to stress $\sigma$. Our energy functional does not include demagnetization nor Zeeman 
terms originating from the dc field  $H_{dc}$ applied along the wire axis. Energy due to the  
ac circular field  $H_{ac}$ created by the injected ac
current $I_{ac}$ is neglected. The anisotropy field as a function of stress is determined by identifying it, 
experimentally, with the value of the dc field $H_{max}$ where the MI curve displays a peak.
Previously Makhnovskiy {\it et al.} \cite{makh} introduced explicitly these terms in the
energy functional with a single helical anisotropy \cite{helical} term in contrast with our work dealing with 
several competing (intrinsic, residual and induced) anisotropies. The stress $\sigma$ (on the surface of the wire)
is related to the torsion angle $\delta$ through:

\begin{equation}
\sigma = (G a \delta)/l
\end{equation}

where $G$ is the shear modulus, $a$ is the radius of the wire and $l$ the length (see fig.2). 
In general, $\sigma$ has a radial dependence since the wire has an inner core with
an axial magnetization surrounded by circular domains \cite{mohri,panina} when  $\lambda^{R}_{s}$ 
is small and negative.
We neglect such dependence and consider, for simplicity, the value of stress on the surface \cite{Raposo} of the wire.
Introducing the saturation magnetization of the samples $M_s$ (measured with a vibrating 
sample magnetometer \cite{BV} as  $\mu_0 M_s =0.8$ Tesla),  the energy may be rewritten as:

\begin{equation}
E= (M_{s}/2 )[H_{K} \sin^{2}(\theta) + H_{\sigma} \sin^{2}(\theta -\delta_T)]
\end{equation}

This shows that we have a competition between a circumferential anisotropy field
$H_{K}=2 K_c/M_s$ and a torsion-stress induced one (called henceforth twist field 
including the residual stress) $H_{\sigma}=3 \lambda_{s}^{R}(\sigma+\sigma_r)/M_s$ . 
Using a Stoner-Wohlfarth approach \cite{SW, landau}, this amounts to
rotate the astroid equation from a set of perpendicular fields $H_x, H_y$:

\begin{equation}
H_x^{2/3}+H_y^{2/3}=H_K^{2/3}
\end{equation}

to another by an angle $\theta^{*}$ resulting in another astroid.
This angle is the same that gives the new equilibrium orientation of the magnetization 
and is simply obtained from the condition minimizing the energy, i.e. 
${[\partial{E}/\partial{\theta}]}_{\theta=\theta^{*}}=0$. We obtain:

\begin{equation}
\tan (2 \theta^{*})= \frac{H_{\sigma}\sin(2\delta_T)}{ H_{K}+ H_{\sigma}\cos(2\delta_T)}
\end{equation}

Using this expression of $\theta^{*}$, the resulting anisotropy field $H^{*}$ is written as:

\begin{equation}
H^{*}= \sqrt{H_{K}^{2}+ H_{\sigma}^{2}+ 2H_{K} H_{\sigma}\cos(2\delta_T)}
\label{field}
\end{equation}

The total energy after this transformation is written as:
\begin{equation}
E= (M_{s}/2 )[H^{*} \sin^{2}(\theta-\theta^{*}) - H^{*} \sin^{2}(\theta^{*}) + H_{\sigma} \sin^{2}(\delta_T)]
\end{equation}

indicating the presence of helical \cite{helical} anisotropy (since we have an angle $\delta_T$ 
in general different from 0 and 90 degrees) 
resulting from the competing circumferential and torsion (applied+residual) induced one. In fact we ought to
find from the MI experiment the angles bounding the value of $\delta_T$.

The MI peak value versus field and torsion angle
should correspond to the anisotropy field $H^{*}$ given by the above formula eq.~\ref{field} 
as explored in the next section.

\section{Numerical procedure}

The MI maximum value field $H_{max}$ is to be fitted to the theoretical expression of $H^{*}$. The
latter depends on four parameters,  $H_{K}$ the zero-stress circumferential 
anisotropy field, $b=3 \lambda_{s}^{R}/M_s$ the normalised magnetostriction constant, 
the angle $\delta_T$ and the residual torsional stress $\sigma_r$. These can be reduced to 
three by normalising all quantities by $H_{K}$.

We use a procedure based on a least squares minimization procedure of the curve
$H^{*}$ versus $\sigma$ to the set of experimental measurements ${[\sigma_{i},  y_{i}]}_{i=1,n}$ where
$y_{i}= H^{*}(\bm{ \alpha}; \sigma_{i})$. $\bm{ \alpha}$ represents the set of parameters
$b,\delta_T$ and $\sigma_r$. The slope of the curve $H^{*}$ versus $\sigma$ that appears to be linear is 
used in the fitting. This choice is sufficient for the description of the method we use, nonetheless 
our algorithm allows us to select any coherent set of criteria we choose to fit the data.

Hence the couple of minimum equations for the data points and the slope are:
\begin{eqnarray}
\frac {1}{n}\sum_{i=1}^{n}[H^{*}(\bm{ \alpha}; \sigma_{i})- y_{i}]^{2} \mbox{  minimum} \\
|\frac{1}{n}\sum_{i=1}^{n}\frac{[b^{2} (\sigma_{i}+\sigma_r) + b H_{K} \cos(2\delta_T)]}{H^{*}(\bm{ \alpha}; \sigma_{i})}- s_{e}|
\mbox{   minimum}
\label{fit}
\end{eqnarray}

where $s_{e}$ is the overall experimental slope.
The  fitting method is based on the Broyden algorithm which is the generalization to
higher dimension of the one-dimensional secant method \cite {NR} and allow us to determine 
in principle two unknowns out of three $b$, $\sigma_r$ and $\delta_T$. Broyden method
is chosen because it can handle underdetermined problems (since we have 2 equations and 3 
unknowns). 

We consider two cases: 
\begin{itemize}
\item Case A: Absence of residual stress ($\sigma_r=0$), then $H_{K}$ is determined as $H_{K}=H^{*}(\bm{ \alpha}; 0)$ and  
$b$, $\delta_T$ are obtained with Broyden algorithm. 
\item Case P: Presence of residual stress, then  $H_{K}$ is a normalising parameter 
and $b$, $\sigma_r$ are evaluated with Broyden algorithm. 
\end{itemize}

The parameter $Ga/l$= 60.82 MPa/radian is obtained directly by averaging over the torsion angles
and the torsion-stress as done in \cite{BV}. Using a wire radius
of $a$=60 $\mu$m and a length of $l$=10 cm, we obtain $G=100$ GPa which is a little different
from the value used in \cite{BV} (60 GPa). 

The fitting shown in figure 3 (corresponding to case P as explained in 
the next section) allows us to predict the anisotropy field $H^{*}$, in 
the negative stress torsion case,
from the experimental results of MI measurements; that is interesting enough, since
the single peak is dominant (at $H_{dc}$=0) and the double peak is barely noticeable in the 
MI curve versus $H_{dc}$. As an example, we can infer that
the anisotropy field $H^{*}$ is zero for a negative value of stress $\sigma \sim$ -150 MPa. 

\section{Results and discussion}
The fitting procedure in case A, discriminates among the positive and negative 
$\lambda_{s}^{R}$ values in favor of the negative value (implying presence of circular
magnetization pattern in the wire). The fitting quality (equivalent
to a $\chi^{2}$ test) is 0.094 when $b$= 1.39 (A/m)/MPa (roughly equal to the slope of $H^{*}$ versus $\sigma$)
the angle $\delta_T$ is almost 180 degrees and $\lambda_{s}^{R}$= 4.68~10$^{-7}$.  
On the other hand, the  fitting quality  is 1.8~10$^{-3}$, with $b$= 1.70 (A/m)/MPa, and $\delta_T \sim 0 \deg$. 
In this case $\lambda_{s}^{R}$=  -4.55~10$^{-7}$ agreeing with the value found by BV
\cite{BV}.

In spite of the successful fit, the result is not acceptable for at least 
two reasons: Firstly, with torsional stress the angle $\delta_T$ should be at least 45 $\deg$
(or between 45$\deg$ and 90 $\deg$ as in Favieres {\it et al.} \cite{Favieres}
and Makhnovskiy {\it et al.} \cite{makh}), 
Secondly the zero-stress anisotropy field  $H_{K}$
($H_{K}=320$ (A/m) \cite{BV} is too large for these wires (For Cobalt rich wires 
$H_{K}$ is on the order of a few tens of A/m. \\
Then we move on to discuss the next case P.
Starting with $H_{K}$=100 A/m, we perform Broyden minimization obtaining: 
$b$=   2.13  (A/m)/MPa yielding  $|\lambda_{s}^{R}|$= 7.13~10$^{-7}$ while $\sigma_r$= 142.4 Mpa.
For  $H_{K}$=50 A/m we obtain: 
$ b$=    2.05  (A/m)/MPa yielding  $|\lambda_{s}^{R}|$= 6.87~10$^{-7}$ while $\sigma_r$= 154.1 Mpa. 
It is remarkable to notice that the fitting curve tapers off for negative stress (see fig. 3) while getting
better for smaller values of $H_{K}$.
For $H_{K}$=25 A/m we obtain:
$ b$=    2.03  (A/m)/MPa yielding  $|\lambda_{s}^{R}|$= 6.83~10$^{-7}$ while $\sigma_r$= 156.5 Mpa.

The value $|\lambda_{s}^{R}|$= 6.83~10$^{-7}$ is slightly larger than found previously by BV.
However, we agree with BV \cite{BV} interpretation of the MI asymmetry in terms of a fabrication induced 
residual torsion stress ($\sim 150$ Mpa) that is
enhanced or counterbalanced (since $H^{*}=0$ for $\sigma=-150$ Mpa) with an applied  positive or 
negative torsional stress.

We conclude that our results are similar to those obtained with the A case and agree as well with
those obtained by BV despite the use in both cases of the linear fit: $H^{*} = H_K + b \sigma$
that allows to extract from the slope of the $H^{*}$ versus $\sigma$ the magnetostriction 
coefficient with the formula \cite{Narita}:

\begin{equation}
\lambda_{s}^{R}=-(\mu_{0} M_{s}/3)(dH^{*}/d\sigma)
\label{lam}
\end{equation}

In order to examine the general case with $\cos(2\delta_T) \ne \pm 1$, we observe that when the magnitude of 
the twist field $H_{\sigma}$ is small with respect to the zero-stress anisotropy field  $H_{K}$
one may expand, to fourth order (for instance), 
the anisotropy field $H^{*}$ with respect to $\sigma$ obtaining: 

\begin{equation}
H^{*}= H_{K} \left\{1 + \frac{H_{\sigma}}{H_{K}} \cos(2\delta_T) + \frac{1}{2} \left[\frac{H_{\sigma}}{H_{K}} \right]^2 \sin^{2}(\delta_T)
         - \frac{1}{2} \left[\frac{H_{\sigma}}{H_{K}}\right]^3 \cos(2\delta_T) \sin^{2}(\delta_T) + O( {\sigma}^4 )\right\}
\label{fieldexp}
\end{equation}

One may also derive a similar expansion for $\lambda_{s}^{R}$ as a function of $\sigma$ \cite{Narita}. 
Using the above relationship eq.~\ref{lam}, one gets:

\begin{equation}
\lambda_{s}^{R}=-(\mu_{0} M_{s} b /3) \left\{ \cos(2\delta_T) + \frac{H_{\sigma}}{H_{K}} \sin^{2}(\delta_T)
         - \frac{3}{2} \left[\frac{H_{\sigma}}{H_{K}}\right]^2 \cos(2\delta_T) \sin^{2}(\delta_T) + O( {\sigma}^3 )\right\}
\label{lamexp}
\end{equation}

Previously, this kind of expansion has been performed by several workers and particularly by
Zhukov {\it et al.} \cite{Zhukov} who made also a fitting of the expansion coefficients as a function of frequency.
This might be included easily in our approach by letting $H_{\sigma}$ vary with frequency.

Generally speaking, the linear regime $H^{*} = H_K \pm b \sigma$ should not be observed when the 
angle $\delta_T$ is different from 0 and 90 $\deg$ ($\pm$ an integer multiple of 90$\deg$).  Then the relationship 
$H^{*}= \sqrt{H_{K}^{2}+ H_{\sigma}^{2}+ 2H_{K} H_{\sigma}\cos(2\delta_T)}$ should apply and when 
the twist field is small ($H_{\sigma} << H_{K}$), the above expressions eq.~\ref{fieldexp},  and eq.~\ref{lamexp} apply.

Experiments versus frequency involving wires with off-wire axis torsion stress conditions ought 
to be done in order to test the general non-linear case (where $H_{\sigma}$ not necessarily being very
small with respect to $H_{K}$), and precision in measuring 
$H^{*}$ versus $\sigma$ should be increased (in the positive torsion angle case of BV~\cite{BV})
 in order to be able to locate accurately the peak of the MI versus field $H_{dc}$ and test the above formula. 
This work is in progress.

\begin{figure}[!h]
\begin{center}
\scalebox{0.6}{\includegraphics[angle=0]{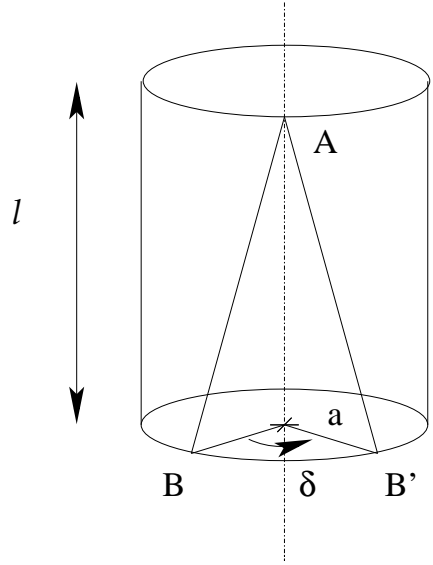}}
\end{center}
\caption{Applying a torsion stress to a wire moves the point $B$ to $B'$ by angle $\delta$.
 The deformation is the displacement per unit length, i.e. $\epsilon=BB'/l=a\delta/l$ where $l$ is the length of the wire.
By Hooke's law, stress and deformation are related by the shear modulus $G$, hence $\sigma = G \epsilon= G (a \delta/l)$}  
\label{fig2}
\end{figure}

\begin{figure}[!h]
\begin{center}
\scalebox{0.5}{\includegraphics[angle=-90]{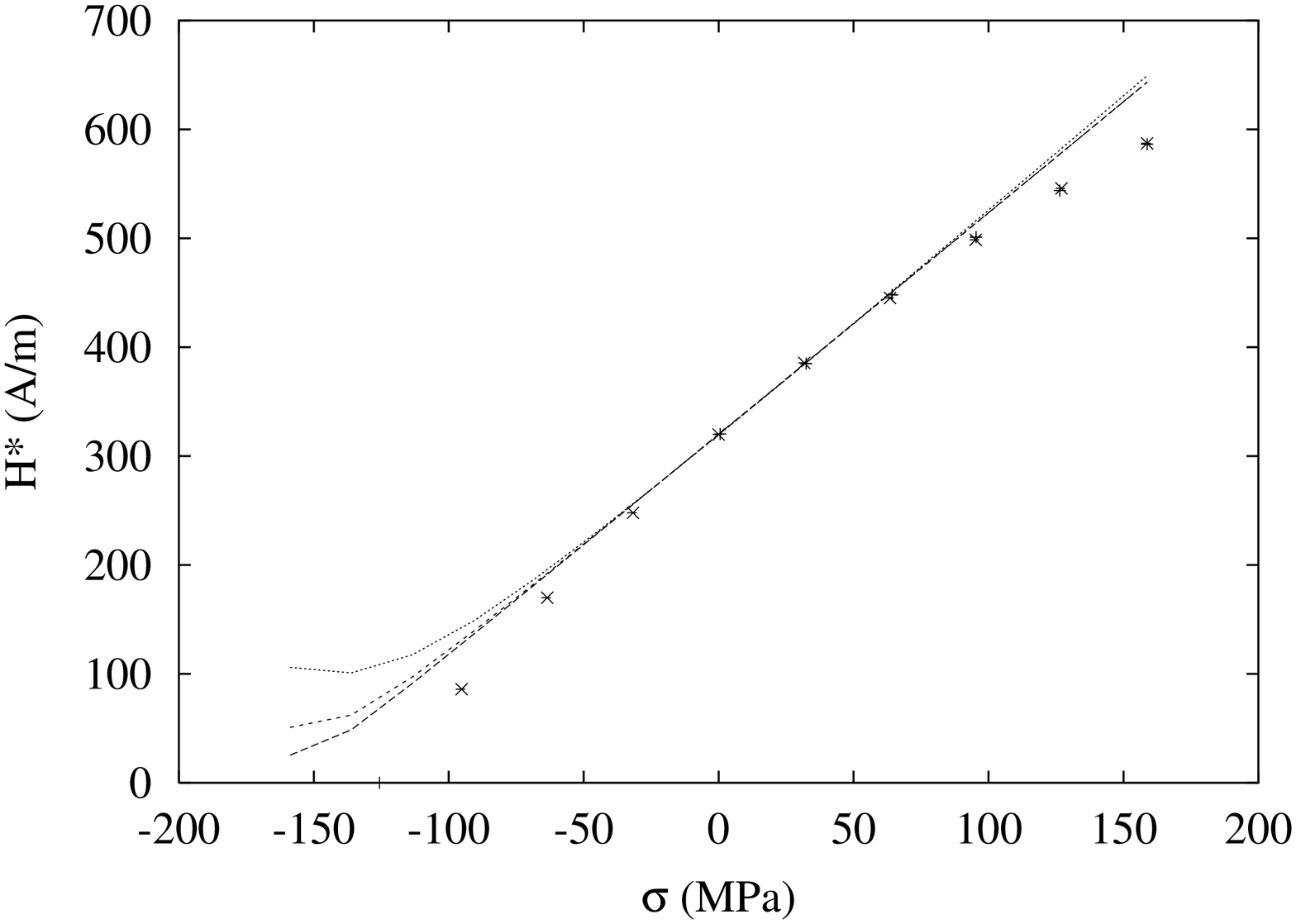}}
\end{center}
  \caption{Experimental results of the anisotropy field $H^{*}$ versus stress as obtained by BV \cite{BV} 
from the field value of $H_{dc}$ at MI peak value and theoretical lines for variable  circular anisotropy
field $H_{K}$ at zero-stress. As $H_{K}$ is decreased from 100 A/m (upper line) to 50 A/m and finally 25 A/m (lowest line) we
get a weaker tapering off of the line for negative stress $\sigma$ and better fitting of the experimental results.
 The residual torsion stress is $\sim 150$ Mpa.}  
\label{fig3}
\end{figure}


\begin{thebibliography}{99}

\bibitem{mohri} K. Mohri, T. Uchiyama, L. P. Shen, C. M. Cai, L. V. Panina, Y. Honkura
and M. Yamamoto,  IEEE Trans. Magn. {\bf 38}  3063 (2002).
\bibitem{panina} L. V. Panina, K. Mohri, T. Uchiyama, M. Noda and K. Bushida, IEEE Trans. Magn. {\bf 31}  1249 (1995).
\bibitem{BV} I. Betancourt and R. Valenzuela, APL {\bf 83}, 2022 (2003).
\bibitem{blanco} J.M. Blanco, A. Zhukov and J. Gonzalez, J. Appl. Phys. {\bf 87} 4813 (2000),
see also: J.M. Blanco, A. Zhukov and J. Gonzalez, J. Phys. D: Appl. Phys. {\bf 32} 3140 (1999).
\bibitem{Ryu} G.H. Ryu, C.S. Yu, C.G. Kim, I.H. Nahm and S. S. Yoon, J. Appl. Phys. {\bf 87} 4828 (2000).
\bibitem{Raposo} V. Raposo, J.M. Gallego and  M. Vazquez  Journal of Magnetism and Magnetic Materials {\bf 242-245} 1435 (2002).
\bibitem{Jiles} M. J. Sablik and D. C. Jiles, IEEE Trans. Magn. {\bf 35}  498 (1999).
\bibitem{Favieres} C. Favieres, C. Aroca M. C. Sanchez and V. Madurga, J. Appl. Phys. {\bf 87} 1889 (2000).
\bibitem{makh} D.P.  Makhnovskiy, L.V. Panina and J. Mapps, Phys. Rev. {\bf B63} (14), 144424 (2001).
\bibitem{helical}  Helical anisotropy is generally accounted for
by introducing an anisotropy axis making an angle $\psi (45\deg  \leq \psi <  90\deg)$
with the wire axis (see for instance ref.~\cite{makh}). Note that torsion stress is also called helical
because it is modeled as tensile and compressive stresses acting perpendicularly on the wire surface 
with an angle of 45 $\deg$ with the wire axis \cite{Jiles}.
\bibitem{constant} The saturation magnetostriction constants in the tensile case ($\lambda_{s}^{T}$) and 
in the torsion case ($\lambda_{s}^{R}$) are, in general, different for a crystalline material. 
In  amorphous materials $\lambda_{s}^{R} \sim \lambda_{s}^{T}$. 
\bibitem{SW} E. C. Stoner and E. P. Wohlfarth,  Philosophical Transactions of the Royal Society
of London {\bf A 240}, 599 (1948).
\bibitem{landau} L. D. Landau and E. M. Lifshitz,  Electrodynamics of Continuous
 Media,  Pergamon,  Oxford,  p.195 (1975).
\bibitem{NR} Numerical Recipes in C: The Art of Scientific Computing, W. H. Press, W. T. Vetterling,
S. A. Teukolsky and B. P. Flannery, Second Edition, page 389, Cambridge University Press (New-York, 1992).
\bibitem{Narita} K. Narita, J. Yamasaki and H. Fukunaga, IEEE Trans. Magn. {\bf 16}, 435 (1980). 
\bibitem{Zhukov} A. Zhukov, V. Zhukova, J.M. Blanco, A.F. Cobe\~{n}o, M.Vazquez and  J. Gonzalez,
Journal of Magnetism and Magnetic Materials  {\bf 258-259} 151 (2003).

 \end{thebibliography}
\end{document}